\documentclass[preprint,letterpaper,english,prl,longbibliography,showpacs,superscriptaddress]{revtex4-1}
\usepackage{color,graphicx,amssymb,amsfonts}
\usepackage[colorlinks=true,linkcolor=blue,citecolor=blue,urlcolor=blue]{hyperref}
\usepackage{rotating,bm,dcolumn,longtable}
\begin{document}

\title{Stability and dopability of native defects and group-V and -VII impurities in single-layer MoS$_2$}

\author{Ji-Young \surname{Noh}}
\affiliation{Korea Research Institute of Standards and Science, Yuseong, Daejeon 305-340, Korea}
\affiliation{Department of Nano Physics, Sookmyung Women's University, Seoul 140-742, Korea}
\author{Minkyu \surname{Park}}
\affiliation{Korea Research Institute of Standards and Science, Yuseong, Daejeon 305-340, Korea}
\affiliation{Department of Nano Science, University of Science and Technology, Daejeon 305-350, Korea}
\author{Yong-Sung \surname{Kim}}
\email{yongsung.kim@kriss.re.kr}
\affiliation{Korea Research Institute of Standards and Science, Yuseong, Daejeon 305-340, Korea}
\affiliation{Department of Nano Science, University of Science and Technology, Daejeon 305-350, Korea}
\author{Hanchul \surname{Kim}}
\affiliation{Department of Nano Physics, Sookmyung Women's University, Seoul 140-742, Korea}
\date{\today}

\begin{abstract}
We investigate the native defects, the Mo substitutional impurities of the group-VB and -VIIB elements, and the S substitutional impurities of the group-VA and -VIIA elements in single-layer MoS$_2$, through density-functional theory calculations.
It is found that the S-vacancy ($V_{\rm S}$) and S-interstitial (S$_i$) are low in formation energy, about $\sim$1 eV, in Mo- and S-rich conditions, respectively, but the carrier doping ability of the $V_{\rm S}$ and S$_i$ is found to be poor, as they are deep level defects.
The V, Nb, and Ta (group-VB) and Re (group-VIIB) impurities are found to be easily incorporated in single-layer MoS$_2$, as Mo substitutional defects, where the V, Nb, and Ta are shallow acceptors and the Re is the only shallow donor among the considered. The unintentional $n$-type doping in single-layer MoS$_2$ exfoliated from the naturally grown MoS$_2$ bulk materials is suggested to originate from the Re impurity.
\end{abstract}


\maketitle

A certain amount of defects and impurities are always present in natural single-crystalline materials.
Even with the small imperfections, the electrical and optical properties of the materials can be largely deviated, and thus the understanding of the role of defects and impurities in the material properties is always important.
The single-layer MoS$_2$ has attracted a great deal of attention recently due to its unique electrical and optical properties \cite{STRANO12}. The single-layer MoS$_2$ can be obtained from the mechanical exfoliation of the naturally grown MoS$_2$ bulk materials \cite{KIS11,NOVOSEL04,NOVOSEL05}, and it shows a direct band gap semiconductor behavior \cite{HEINZ10,WANG10,KLEIN01} with typically unintensionally $n$-type doped electrical conductivity \cite{KIS11,NOVOSEL05}.

The fundamental band gap of a single-layer MoS$_2$ is 2.8 eV \cite{RAMASU12,LAMBRE12}. With the large exciton binding energy of about 0.9 eV, the optical band gap has been measured at 1.9 eV from the absorption and photo-luminescence spectroscopy \cite{HEINZ10,WANG10}. In $n$-type single-layer MoS$_2$ thin film transistors, the optically excited trions comprising two electrons and one hole have been also identified in presence of carrier electrons \cite{SHAN13}.
Besides the large electronic correlation effects, the intrinsic broken inversion symmetry of the single-layer MoS$_2$ induces the spin-orbit split bands at K and -K points in the hexagonal Brillouin zone \cite{SCHWINGEN11} and opens the new era of the spin- and valley-controlled electronics \cite{FENG12,CUI12,HEINZ12,LIU12,XIAO12}.

The practical application of the single-layer MoS$_2$ is not trivial as well. The electron carrier mobility of the single-layer MoS$_2$ has been measured to be as large as $\sim$200 cm$^2$V$^{-1}$s$^{-1}$ in the HfO$_2$/MoS$_2$/SiO$_2$ stacked thin film transistors \cite{KIS11}, while it is quite smaller, about $\sim$1 cm$^2$V$^{-1}$s$^{-1}$, in the air/MoS$_2$/SiO$_2$ stacked structure \cite{NOVOSEL05}. Although the role of the encapsulating dielectric layers is not clear yet, the measured large electron mobility makes the single-layer MoS$_2$ promising for the next generation high-speed electronics.
For the electronics application of the single-layer MoS$_2$, the efficient and controllable $n$- and $p$-type doping technology is an essential constituent.

In this study, we investigate the stabilities and the electrical dopabilities of the native defects of S-vacancy ($V_{\rm S}$), S-interstitial (S$_i$), Mo-vacancy ($V_{\rm Mo}$), and Mo-interstitial (Mo$_i$); the Mo substitutional impurities of the group-VB (V, Nb, Ta) and the group-VIIB (Mn, Tc, Re) elements; and the S substitutional impurities of the group-VA (N, P, As, Sb) and the group-VIIA (F, Cl, Br, I) elements in single-layer MoS$_2$, through density-functional theory (DFT) calculations.
For the native defects, the $V_{\rm S}$ and S$_i$ are found to be low in formation energy, about $\sim$1 eV, in Mo- and S-rich conditions, respectively, while the Mo-related defects of the $V_{\rm Mo}$ and Mo$_i$ have very high formation energies of above 2.5 eV. The carrier doping ability of the $V_{\rm S}$ and S$_i$ is however found to be poor, as both the $V_{\rm S}$ and S$_i$ are deep level defects.
The group-VB impurities of V, Nb, and Ta, and the group-VIIB Re impurity as Mo substitutional defects are found to be negative in formation energy in S-rich conditions, by which their incorporation in single-layer MoS$_2$ is facile.
The V, Nb, and Ta are found to be shallow acceptors, and the Re is the only shallow donor impurity among the considered defects.
Since Re is well known to exist in naturally grown MoS$_2$ (molybdenite) via geochemical reactions since $\sim$2.91 gigayears ago (Ga) in the oxygenized earth subsurfaces \cite{HAZEN13}, the unintentional $n$-type doping in single-layer MoS$_2$ exfoliated from it is suggested to originate from the Re impurity.

\section*{Results}
The calculated formation energies of the native defects and impurities are listed in Table \ref{table1}.
Among the native defects, the S-related defects of $V_{\rm S}$ and S$_i$ are found to be low in formation energy. In Mo-rich limit condition, the most stable native defect is $V_{\rm S}$, of which formation energy is in the range of 0.70-1.48 eV depending on the Fermi level: 1.22 eV, when the Fermi level is at the VBM, and 0.70 eV, when the Fermi level is at the CBM.
In S-rich limit condition, the S$_i$ is the most stable native defect, of which formation energy is in the range of 0.72-0.99 eV depending on the Fermi level.
On the other hand, the Mo-related native defects are found to have very high formation energies above 2.5 eV, which indicates that the $V_{\rm Mo}$ and Mo$_i$ defects are rare in single-layer MoS$_2$.

\begin{figure}[t!]
\includegraphics[angle=0,width=0.5\linewidth]{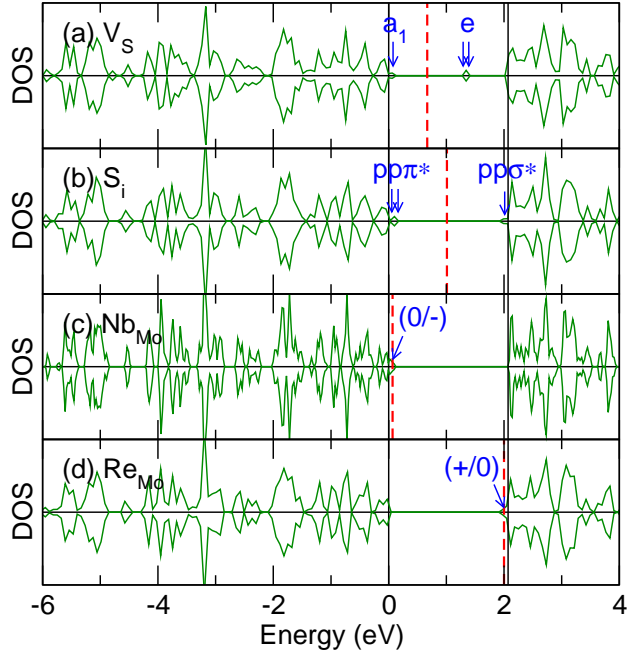}\\
\caption{{\bf Density-of-states of the native defects and impurities.} Electronic density-of-states (DOS) of the (a) V$_{\rm S}$, (b) S$_i$, (c) Nb$_{\rm Mo}$, and (d) Re$_{\rm Mo}$ in the neutral charge state. The (red) dashed vertical lines indicate the Fermi level, and the (blue) arrows indicate the defect states inside the band gap.}
\label{fig1}
\end{figure}

\begin{figure}[t!]
\includegraphics[angle=0,width=0.5\linewidth]{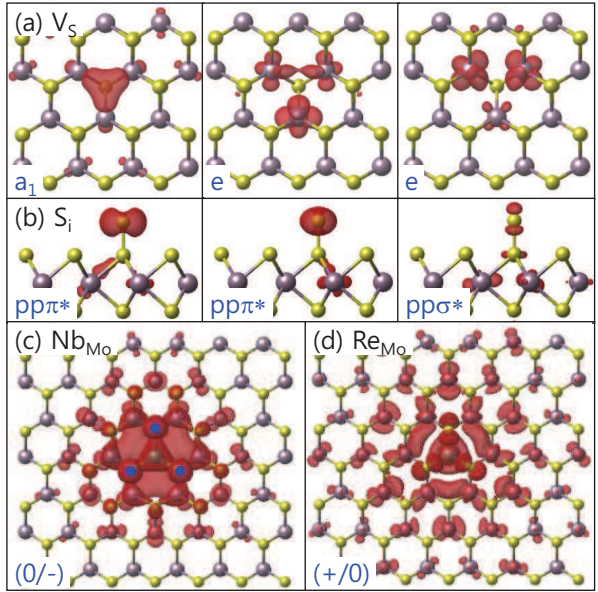}\\
\caption{{\bf Atomic structures of the native defects and charge densities of the defect states.} Stable atomic configurations of (a) $V_{\rm S}$, (b) S$_i$, (c) Nb$_{\rm Mo}$, and (d) Re$_{\rm Mo}$, and the charge densities for the defect states are shown. The charge isosurfaces are 3.83$\times$10$^{-7}$\AA$^{-3}$ for (a) and (b), and 3.83$\times$10$^{-8}$\AA$^{-3}$ for (c) and (d). }
\label{fig2}
\end{figure}

Thus, the abundant native defects in single-layer MoS$_2$ are $V_{\rm S}$ and S$_i$ in equilibrium, of which one is dominant is depending on the stoichiometry.
The electronic density-of-states (DOS) of the $V_{\rm S}$ and S$_i$ in the neutral charge state are shown in Fig. \ref{fig1}.
For the $V_{\rm S}$, three defect states are induced inside the band gap. Since the $V_{\rm S}$ has trigonal symmetry [Fig. \ref{fig2}(a)], the defect states are classified as a singlet $a_1$ state and doubly degenerate $e$ states. Due to the absence of an anion ($V_{\rm S}$) in MoS$_2$, two excess electrons occupy the $a_1$ state, and the two $e$ states are empty in the neutral charge state. The $a_1$ level is found to be located near the VBM, and thus it is not easy to generate carrier electrons from the $V_{\rm S}^0$. Instead, the empty $e$ levels are inside the band gap, as shown in Fig. \ref{fig1}(a), and thus the $V_{\rm S}$ can be an electron trap center in single-layer MoS$_2$. When the Fermi level is above the (0/-) transition level of $\epsilon_V$+1.304 eV, the $V_{\rm S}$ can trap an electron with ionized into $V_{\rm S}^-$, and when the Fermi level is above the (-/2-) transition level of $\epsilon_V$+1.872 eV, the $V_{\rm S}$ is ionizied into $V_{\rm S}^{2-}$.
When the Fermi level is near the VBM below the (+/0) transition level of $\epsilon_V$+0.261 eV, the $V_{\rm S}$ also can trap one hole with ionized into $V_{\rm S}^+$.

The S$_i$ is found to be also a deep level defect. When the Fermi level is near the VBM below the (+/0) transition level of $\epsilon_V$+0.267 eV, the S$_i$ traps a hole (S$_i^+$), and when the Fermi level is near the CBM above the (0/-) transition level of $\epsilon_V$+1.813 eV, the S$_i$ traps an electron (S$_i^-$). In a wide range of the Fermi level between the (+/0) and (0/-) transition levels, the S$_i$ exists in the neutral charge state (S$_i^0$).
The atomic structure of the S$_i$ is shown in Fig. \ref{fig2}(b), in which the S$_i$ is located on top of a S atom on the single-layer MoS$_2$. The S on-top configuration is found to be the lowest energy configuration for S$_i$, where the configurations of the bridge between two S atoms and the hexagonal interstitial in the Mo layer are found to be much higher in energy than the S on-top configuration by 2.8 and 6.4 eV, respectively, in the neutral charge state.
In the S on-top configuration, the S$_i$ forms a S-S bond with a host S atom [Fig. \ref{fig2}(b)]. As shown in Fig. \ref{fig1}(b), three defect states are generated by the S$_i$; degenerated two are occupied $p_xp_x\pi^*$ and $p_yp_y\pi^*$ states near the VBM, and one is an empty $p_zp_z\sigma^*$ state near the CBM [see Fig. \ref{fig2}(b)].
This electronic configuration of the S$^{2-}$-S$_i^0$ is very similar to the peroxide (O$_2^{2-}$) configuration \cite{NAHM12}.
Since the level difference between the $pp\pi^*$ and $pp\sigma^*$ is large, neither the excitation of the $pp\pi^*$ electrons nor the occupation of the $pp\sigma^*$ level is unlikely in single-layer MoS$_2$, unless the Fermi level is very close to the VBM or CBM.
Thus, the S$_i$ is only a deep level defect without giving an efficient carrier doping in single-layer MoS$_2$ even in high concentration of S$_i$.

Where the carrier doping ability of the native defects in single-layer MoS$_2$ is found to be poor, the impurities, which can be introduced in MoS$_2$ either intentionally or naturally, are investigated.
We consider the group-VA (N, P, As, Sb) and the group-VIIA (F, Cl, Br, I) elements as S substitutional impurities, and the group-VB (V, Nb, Ta) and the group-VIIB (Mn, Tc, Re) elements as Mo substitutional impurities.
The formation energies and the electron transition levels are summarized in Table \ref{table1}.

For acceptors, when the Fermi level is close to the VBM, the S substitutional N$_{\rm S}$, P$_{\rm S}$, As$_{\rm S}$, and Sb$_{\rm S}$ impurities are unlikely to be incorporated in single-layer MoS$_2$ in S-rich conditions by their high positive formation energies (see Table \ref{table1}), which indicates that the N$_2$ molecule, P triclinic metal, As$_2$ molecule, and Sb trigonal metal forms are more stable than the doped configurations, respectively.
In Mo-rich conditions, only when the Fermi level is near the CBM, the formation energies of the acceptor-like defects are found to be negative, and thus they can act as electron traps in the $n$-type MoS$_2$. When the Fermi level is close to the VBM, the formation energies of the N$_{\rm S}$, P$_{\rm S}$, As$_{\rm S}$, and Sb$_{\rm S}$ are found to be positive even in the Mo-rich conditions, and thus their concentrations cannot be high in equilibrium. Thus, the $p$-type doping ability of the group-VA impurities is low in single-layer MoS$_2$.
On the other hand, the group-VB impurities at the Mo substitutional sites have negative formation energies in S-rich conditions even when the Fermi level is close to the VBM, which indicates that the V$_{\rm Mo}$, Nb$_{\rm Mo}$, and Ta$_{\rm Mo}$ acceptors are easily incorporated in the $p$-type single-layer MoS$_2$. The (0/-) acceptor transition levels are found to be shallow for all the group-VB impurities, which are only 61, 29, 46 meV above the VBM for the V$_{\rm Mo}$, Nb$_{\rm Mo}$, and Ta$_{\rm Mo}$, respectively. [For Nb$_{\rm Mo}$, the calculated DOS is shown in Fig. \ref{fig1}(c) and the charge density of the shallow acceptor level is shown in Fig. \ref{fig2}(c).] Therefore, we suggest that the V, Nb, and Ta impurities are efficient $p$-type dopants in single-layer MoS$_2$.

For donors, the S substitutional F$_{\rm S}$, Cl$_{\rm S}$, Br$_{\rm S}$, and I$_{\rm S}$ impurities can be incorporated in $p$-type single-layer MoS$_2$ in Mo-rich conditions, as hole compensating centers, with their negative formation energies (see Table \ref{table1}). In $n$-type single-layer MoS$_2$, only the F$_{\rm S}$ and Cl$_{\rm S}$ can be incorporated in Mo-rich conditions, as possible donors. In S-rich conditions, the formation energies are even higher.
In any ways, all the S substitutional group-VIIA impurities are found to be deep donors with the (+/0) transition levels of 387, 160, 260, and 331 meV below the CBM for the F$_{\rm S}$, Cl$_{\rm S}$, Br$_{\rm S}$, and I$_{\rm S}$, respectively.
Based on their relatively high formation energies (except F$_{\rm S}$) and their deep donor characters, the F$_{\rm S}$, Cl$_{\rm S}$, Br$_{\rm S}$, and I$_{\rm S}$ are concluded to be not efficient for $n$-type doping in single-layer MoS$_2$.
On the other hand, the donor-like Mo substitutional group-VIIB impurities of the Mn$_{\rm Mo}$, Tc$_{\rm Mo}$, and Re$_{\rm Mo}$ are found to be easily incorporated in $n$-type single-layer MoS$_2$, as they have negative formation energies in S-rich conditions (see Table \ref{table1}).
The (+/0) donor transition levels of the Mn$_{\rm Mo}$, Tc$_{\rm Mo}$, and Re$_{\rm Mo}$ are found to be 660, 208, and 61 meV below the CBM, respectively. The Re$_{\rm Mo}$ is found to be the only shallow donor impurity among them, and thus the Re impurity can be an efficient $n$-type dopant in single-layer MoS$_2$.
[For Re$_{\rm Mo}$, the calculated DOS is shown in Fig. \ref{fig1}(d) and the charge density of the shallow donor level is shown in Fig. \ref{fig2}(d).]

The presence of Re in naturally grown MoS$_2$ bulk materials (molybdenite) has been well known in earth science. By the ``Great Oxygenation Event'' (GOE), the concentration of atmospheric oxygen had been increased during the Paleoproterozoic Era ($\sim$2.5-1.6 Ga), and it had been furthermore increased following episodes of Neoproterozoic Era ($\sim$1.0-0.541 Ga) glaciation \cite{KUMP08}. Both Mo and Re are insoluble in their more reduced tetravalent forms: Mo$^{2+}$ and Re$^{3+}$, whereas Mo$^{6+}$ and Re$^{7+}$ are more soluble in aqueous solutions and thus mobilized under more oxidizing conditions \cite{MILLER11}. The association of MoO$_4^{2-}$, ReO$_4^{-}$, and SO$_4^{2-}$ in the dissolved phase and that of Mo, Re, and S as sulfides in the crust have been observed globally \cite{MILLER11}.
The Re contents in molybdenite show overall increase from 832 ppm (1.63$\times$10$^{19}$ cm$^{-3}$) to 10424 ppm (2.04$\times$10$^{20}$ cm$^{-3}$) through time when the molybdenite was formed since $\sim$2.91 Ga \cite{HAZEN13}.
Thus, the unintentional $n$-type doping in single-layer MoS$_2$ exfoliated from it is suggested to be due to the Re impurity.

In conclusion, the $V_{\rm S}$ and S$_i$ native defects are suggested to be low in formation energy in Mo-rich and S-rich conditions, respectively, in single-layer MoS$_2$, and their concentrations are expected to be high in thermal equilibrium. However, their carrier doping ability in single-layer MoS$_2$ is not found to be strong, where the $V_{\rm S}$ and S$_i$ act as deep trap centers for electrons and holes.
The S substitutional group-VA and -VIIA impurities are neither found to be efficient $p$- and $n$-type dopants, because they have high formation energies to incorporate and/or deep level defect characters. It is suggested that the V$_{\rm Mo}$, Nb$_{\rm Mo}$, and Ta$_{\rm Mo}$ impurities can be efficient $p$-type dopants, since they have low formation energies and are shallow acceptors. The Re$_{\rm Mo}$ is found to be the only shallow donor impurity among the considered and easily incorporated. For intentional $n$-type doping in single-layer MoS$_2$, the Re is suggested to be an efficient $n$-type dopant. In naturally grown MoS$_2$, a certain amount of Re impurities are known to exist, and then the single-layer MoS$_2$ exfoliated from it can have the $n$-type conductivity unintentionally by the naturally associated Re impurities.

\section*{Methods}
We performed DFT calculations as implemented in the Vienna ab initio simulation package (VASP) code \cite{VASP1,VASP2}. The projector augmented wave (PAW) pseudopotentials \cite{PAW1,PAW2} and the local density approximation (LDA) \cite{LDA} for the exchange-correlation energy of electrons were employed.
The 8$\times$8 supercell with 192 host atoms with the vacuum region of 18 \AA\ was used for the single-layer MoS$_2$, where the native defects and impurities were incorporated.
A kinetic energy cutoff of 350 eV for the plane-wave expansion and a single $k$-point at $\Gamma$ for the Brillouin zone integration were used. The atomic structures were relaxed until the Hellmann-Feynman forces were less than 0.02 eV/\AA.
The formation energies ($E_f$) of the native defects and impurities were calculated from
\begin{equation}
E_f(D^q)=E_t(D^q)-\sum_iN_i\mu_i+q(\epsilon_V+\epsilon_f),
\label{eq1}
\end{equation}
where $D^q$ is the defect with the charge state $q$, $E_t(D^q)$ is the total energy of the supercell containing the defect $D^q$, $N_i$ is the number of element $i$ in the supercell (Mo, S, and impurity),
$\mu_i$ is the chemical potential of the element $i$, $\epsilon_V$ is the valence band maximum (VBM) level of the perfect single-layer MoS$_2$, and $\epsilon_f$ is the Fermi level with reference to the VBM.
The maximum $\epsilon_f$ we considered is the calculated LDA band gap of 2.06 eV, which corresponds to the conduction band minimum (CBM) level.
For charged defects, the total energies were corrected to eliminate the spurious Madelung interactions between the supercells by -0.35$\times$$\mid$$q$$\mid$ eV, obtained for our supercell structures.
The $\mu_\mathrm{Mo}$ and $\mu_\mathrm{S}$ are correlated by $\mu_\mathrm{Mo}+2\times\mu_\mathrm{S}=\mu_\mathrm{MoS_2}$, where $\mu_\mathrm{MoS_2}$ is the total energy of the perfect single-layer MoS$_2$ per a formula unit.
For the Mo-rich limit condition, $\mu_{\mathrm{Mo}, max}$$=$$E_t$(Mo-metal), where $E_t$(Mo-metal) is the DFT total energy of a Mo bcc metal per an atom.
Then, $\mu_{\mathrm{S}, min}=(\mu_\mathrm{MoS_2}-\mu_{\mathrm{Mo}, max})/2$.
The heat of formation of a single-layer MoS$_2$ calculated from $H_\mathrm{MoS_2}=\mu_{\mathrm{S}, max}-\mu_{\mathrm{S}, min}$ is 1.44 eV. We set $\mu_{\mathrm{S}, max}=0$ eV for convention, and then the $\mu_\mathrm{S}$ range for stable single-layer MoS$_2$ formation is -1.44 eV (Mo-rich limit) $<$ $\mu_\mathrm{S}$ $<$ 0 eV (S-rich limit).
The chemical potentials for the impurity atoms of V, Nb, Ta; Mn, Tc, Re; N, P, As, Sb; F, Cl, Br, I were  obtained from the DFT total energies of a V bcc metal, a Nb bcc metal, a Ta bcc metal, an $\alpha$-Mn monoclinic metal, a Tc hcp metal, a Re hcp metal, a N$_2$ molecule, a P triclinic metal, an As$_2$ molecule, an Sb trigonal metal, a F$_2$ molecule, a Cl$_2$ molecule, a Br$_2$ molecule, and an I$_2$ molecule, respectively, per an atom.

\newcommand{\NNT}{\it Nature Nanotech. }
\newcommand{\NMAT}{\it Nature Mater. }
\newcommand{\NCOM}{\it Nature Comm. }
\newcommand{\SCI}{\it Science }
\newcommand{\NL}{\it Nano Lett. }
\newcommand{\PNAS}{\it Proc. Natl Acad. Sci. USA }
\newcommand{\PRL}{\it Phys. Rev. Lett. }
\newcommand{\APL}{\it Appl. Phys. Lett. }
\newcommand{\CRL}{Chem. Phys. Lett. }
\newcommand{\PR}{\it Phys. Rev. }
\newcommand{\JAP}{\it J. Appl. Phys. }
\newcommand{\PRB}{\it Phys. Rev. B }
\newcommand{\SSC}{\it Solid State Commun. }
\newcommand{\PSS}{\it Phys. Solid State }
\newcommand{\PSSB}{\it Phys. Status Solidi B }
\newcommand{\JJAP}{\it Jpn. J. Appl. Phys. }
\newcommand{\JPA}{\it J. Phys. A }
\newcommand{\JPCB}{\it J. Phys. Chem. B }
\newcommand{\JPC}{\it J. Phys. Chem. }
\newcommand{\JCG}{\it J. Cryst. Growth }

\section*{Acknowledgments}
\begin{acknowledgments}
This work was supported by Nano R\&D Program through the National Research Foundation (NRF) of Korea (No. 2013-0042633).
\end{acknowledgments}

\section*{Author contribution}
J.N. performed DFT calculations for native-defects, and M.P. performed DFT calculations for impurities.
Y.K. conducted the research. J.N., M.P., and Y.K. wrote the manuscript. Y.K. is responsible for coordinating the project.

\section*{Additional information}
{\bf Competing financial interests:} The authors declare no competing financial interests.

\newpage

\begin{table}[t]
\tiny
\begin{ruledtabular}
\begin{tabular}{lcccccl}
Groups          & Defects       & \multicolumn{4}{c}{Formation energies (eV)}                                         & Transition levels (eV) \\
                &               & \multicolumn{2}{c}{Mo-rich limit}       & \multicolumn{2}{c}{S-rich limit}          &                        \\
                &               & VBM               & CBM                 & VBM                 & CBM                 &                        \\ \hline
Native defects  &               &                   &                     &                     &                     &                        \\
                & $V_{\rm S}$   & 1.48 (0)          & 0.81 (-)            & 2.92 (0)            & 2.25 (-)            & $\epsilon_V$+0.261 (+/0) \\
                &               & 1.22 (+)          & 0.70 (2-)           & 2.66 (+)            & 2.14 (2-)           & $\epsilon_V$+1.304 (0/-) \\
                &               &                   &                     &                     &                     & $\epsilon_V$+1.872 (-/2-)\\
                & S$_i$ (on-top)& 2.43 (0)          & 2.43 (0)            & 0.99 (0)            & 0.99 (0)            & $\epsilon_V$+0.267 (+/0) \\
                &               & 2.16 (+)          & 2.26 (-)            & 0.72 (+)            & 0.82 (-)            & $\epsilon_V$+1.813 (0/-) \\
                & $V_{\rm Mo}$  & 7.18 (+)          & 6.50 (-)            & 4.30 (+)            & 3.62 (-)            & $\epsilon_V$+0.652 (+/-) \\
                &               &                   & 5.70 (2-)           &                     & 2.82 (2-)           & $\epsilon_V$+1.178 (-/2-)\\
                &               &                   & 5.46 (3-)           &                     & 2.58 (3-)           & $\epsilon_V$+1.740 (2-/3-)\\
                & Mo$_i$ (split)& 3.36 (+)          & 4.27 (0)            & 6.24 (+)            & 7.15 (0)            & $\epsilon_V$+0.370 (2+/+)\\
                &               & 2.99 (2+)         & 3.97 (-)            & 5.87 (2+)           & 6.85 (-)            & $\epsilon_V$+0.904 (+/0) \\
                &               &                   &                     &                     &                     & $\epsilon_V$+1.686 (0/-) \\
Group-VA impurities &           &                   &                     &                     &                     &                          \\
                & N$_{\rm S}$   & 0.76 (-)          & -1.40 (-)           & 2.20 (-)            & 0.03 (-)            & $\epsilon_V$+0.455 (0/-) \\
                &               & 0.31 (0)          &                     & 1.74 (0)            &                     &                          \\
                & P$_{\rm S}$   & 0.47 (-)          & -1.70 (-)           & 1.91 (-)            & -0.26 (-)           & $\epsilon_V$+0.137 (0/-) \\
                &               & 0.33 (0)          &                     & 1.77 (0)            &                     &                          \\
                & As$_{\rm S}$  & 0.80 (-)          & -1.37 (-)           & 2.24 (-)            & 0.07 (-)            & $\epsilon_V$+0.087 (0/-) \\
                &               & 0.71 (0)          &                     & 2.15 (0)            &                     &                          \\
                & Sb$_{\rm S}$  & 1.44 (-)          & -0.72 (-)           & 2.88 (-)            & 0.72 (-)            & $\epsilon_V$+0.341 (0/-) \\
                &               & 1.10 (0)          &                     & 2.54 (0)            &                     &                          \\
Group-VB impurities &           &                   &                     &                     &                     &                          \\
                & V$_{\rm Mo}$  & 0.61 (-)          & -1.56 (-)           & -2.27 (-)           & -4.44 (-)           & $\epsilon_V$+0.061 (0/-) \\
                &               & 0.55 (0)          &                     & -2.33 (0)           &                     &                          \\
                & Nb$_{\rm Mo}$ & 0.18 (-)          & -1.98 (-)           & -2.69 (-)           & -4.86 (-)           & $\epsilon_V$+0.029 (0/-) \\
                &               & 0.15 (0)          &                     & -2.72 (0)           &                     &                          \\
                & Ta$_{\rm Mo}$ & 0.08 (-)          & -2.08 (-)           & -2.79 (-)           & -4.96 (-)           & $\epsilon_V$+0.046 (0/-) \\
                &               & 0.04 (0)          &                     & -2.84 (0)           &                     &                          \\
Group-VIIA impurities &         &                   &                     &                     &                     &                          \\
                & F$_{\rm S}$   & -3.62 (+)         & -1.46 (+)           & -2.18 (+)           & -0.02 (+)           & $\epsilon_C$-0.387 (+/0) \\
                &               &                   & -1.85 (0)           &                     & -0.41 (0)           &                          \\
                & Cl$_{\rm S}$  & -2.27 (+)         & -0.10 (+)           & -0.83 (+)           & 1.34 (+)            & $\epsilon_C$-0.160 (+/0) \\
                &               &                   & -0.26 (0)           &                     & 1.18 (0)            &                          \\
                & Br$_{\rm S}$  & -1.86 (+)         & 0.31 (+)            & -0.42 (+)           & 1.75 (+)            & $\epsilon_C$-0.260 (+/0) \\
                &               &                   & 0.05 (0)            &                     & 1.49 (0)            &                          \\
                & I$_{\rm S}$   & -1.42 (+)         & 0.74 (+)            & 0.02 (+)            & 2.18 (+)            & $\epsilon_C$-0.331 (+/0) \\
                &               &                   & 0.41 (0)            &                     & 1.85 (0)            &                          \\
Group-VIIB impurities &         &                   &                     &                     &                     &                          \\
                & Mn$_{\rm Mo}$ & 0.61 (+)          & 2.77 (+)            & -2.27 (+)           & -0.11 (+)           & $\epsilon_C$-0.660 (+/0) \\
                &               &                   & 2.11 (0)            &                     & -0.76 (0)           &                          \\
                & Tc$_{\rm Mo}$ & -0.60 (+)         & 1.57 (+)            & -3.47 (+)           & -1.31 (+)           & $\epsilon_C$-0.208 (+/0) \\
                &               &                   & 1.36 (0)            &                     & -1.52 (0)           &                          \\
                & Re$_{\rm Mo}$ & -0.61 (+)         & 1.56 (+)            & -3.49 (+)           & -1.32 (+)           & $\epsilon_C$-0.061 (+/0) \\
                &               &                   & 1.49 (0)            &                     & -1.38 (0)           &                          \\
\end{tabular}
\end{ruledtabular}
\caption{{\bf Formation energies and transition levels of the native defects and impurities.} Formation energies of the native defects and the group-V and -VII impurities in single-layer MoS$_2$ in the Mo-rich and S-rich limit conditions are listed. The formation energies of the defects in general stoichiometry conditions that the single-layer MoS$_2$ is stable are in between the two limits.
The Fermi levels at the VBM and the CBM are considered, and the charge states of the defects are indicated in the parenthesis. The transition levels between two charge states that are inside the band gap are listed. The $\epsilon_V$ indicates the VBM level, and the $\epsilon_C$ is the CBM level.}\label{table1}
\end{table}

\end{document}